\documentclass[runningheads]{svmult}

\usepackage{makeidx}   
\usepackage{graphicx}  
\usepackage{multicol}  
\usepackage{physprbb}  
\makeindex             

\begin{document}
\title*{Probes of Jet-Disk-Coupling in AGN from combined 
VLBI and X-Ray Observations}
\toctitle{Probes of Jet-Disk-Coupling in AGN from combined
VLBI and X-Ray Observations}
%
%
\titlerunning{Probes of Jet-Disk-Coupling in AGN from combined VLBI and X-Ray Obs.}
%
\author{M.\,Kadler\inst{1}
\and J.\,Kerp\inst{2}
\and E.\,Ros\inst{1}
\and K.\,A.\,Weaver\inst{3}
\and J.\,A.\,Zensus\inst{1}}
\authorrunning{Matthias Kadler et al.}
%
%
\institute{Max-Planck-Institut f\"ur Radioastronomie,
Auf dem H\"ugel 69, D-53121 Bonn, Germany
\and
Radioastronomisches Institut, Universit\"at Bonn,
Auf dem H\"ugel 71, D-53121 Bonn, Germany
\and
Laboratory for High Energy Astrophysics, NASA/Goddard Space Flight Center, Greenbelt, MD 20771, U.S.A. 
}

\maketitle              

\begin{abstract}
The formation of powerful extragalactic jets is not well understood at present
as well as the associated key question:``What makes an AGN radio loud?''.
Here we discuss how the combination of VLBI- and X-ray
spectroscopic observations allows the inter-relation between the accretion flow
and the formation of relativistic jets in AGN to be explored.
\end{abstract}

\paragraph{\bf The Present: }
The nearby, radio-loud, core-dominated active galaxy NGC\,1052 exhibits
strong, relativistically broadened iron-line emission at X-ray frequencies
(Kadler et al. 2004).
Pronounced variability of the broad iron line was accompanied by
an ejection of relativistic plasma into the radio jet in early 2000.
This behaviour can be interpreted 
as an instability 
of the inner accretion disk around the central supermassive black hole, which
triggered enhanced accretion 
onto the black hole, while a fraction of the inner-disk material
was injected into the jet. 
This observational finding demonstrates that 
the combined analysis of VLBI- and X-ray spectroscopic data 
allows the interplay of black hole accretion dynamics and jet production
in active galaxies to be studied directly.

\paragraph{\bf The Future: }
NGC\,1052 is the only AGN with strong broad iron line emission
and a bright, compact radio jet
detected so far. This makes NGC\,1052 the key object
for
future studies, which address the open questions of black hole accretion and
jet formation. A dedicated monitoring campaign of NGC\,1052
with {\sc i)} high frequency VLBI observations and {\sc ii)} high
signal-to-noise X-ray spectroscopic observations provides the ideal approach
towards an understanding of the unsolved questions in AGN physics.

The performance of VLBI observations is being continuously improved towards
higher frequencies (yielding higher angular resolution) and higher sensitivities.
Particularly, an oncoming Square Kilometer Array (SKA),
capable of operating at high frequencies with long baselines, will
provide a breakthrough in sensitivity. This will allow a much larger number
of AGNs, particularly some radio-quiet (i.e., weak) objects, to be accessed.

A significant improvement of black-hole accretion and
jet formation studies in AGN will come from future X-ray observatories.
In Fig.~1, we show simulated NGC\,1052 X-ray spectra of current and future X-ray
telescopes in comparison to a measured {\it XMM-Newton} spectrum
(Kadler et al. 2004). 
{\it ASTRO-E2} (Hajime 2003), which will be launched in early 2005,
will yield high X-ray spectral resolution throughout the
iron line emission energy range.
With {\it Constellation-X} (White \& Tananbaum 2003) and {\it XEUS} (Parmar et al. 2004) the sensitivity of broad iron line studies
will increase dramatically.
This will disclose other radio-loud AGN with considerable broad iron
line emission. Direct comparison of the accretion flow
properties in
radio-loud and radio-quiet AGN will allow the unsolved mystery of
the radio loud/quiet phenomenon
to be attacked, or in other words to address the question:
``What leads some supermassive black holes to launch powerful
relativistic jets?''

\begin{figure}[b]
\begin{center}
\includegraphics[width=0.8\textwidth]{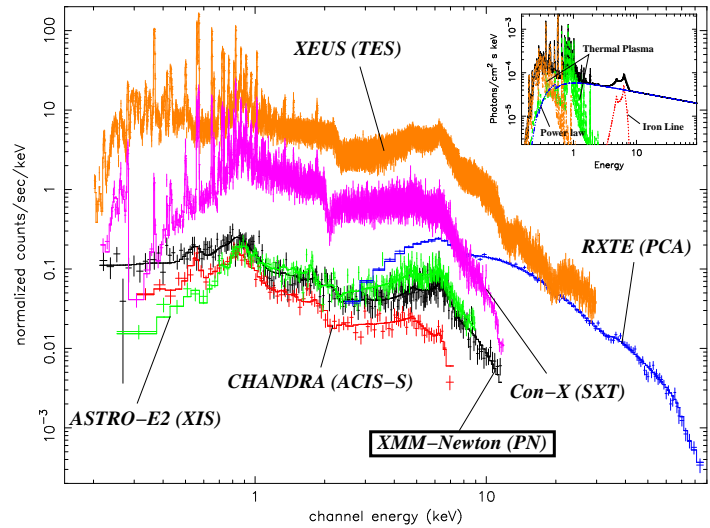}
\end{center}
\caption[]{\scriptsize The measured 13\,ksec
{\it XMM-Newton} spectrum of NGC\,1052 (Kadler et al. 2004) 
and simulated 13\,ksec spectra from {\it CHANDRA},
{\it RXTE}, {\it ASTRO-E2}, {\it XEUS}, and {\it Con-X}. The spectral model
consisting of absorbed plasma, power-law, and iron-line
emission is shown in the inset panel.}
\label{eps1}
\end{figure}

%


\begin{thebibliography}{8.}
\addcontentsline{toc}{section}{References}

\bibitem{Kad04} Kadler, M., Ros, E., Weaver, K., A., Kerp, J., Zensus, J. A. 2004,
BAAS Vol. 36, No.\,2, 823

\bibitem{Haj03} Hajime, I. 2003, X-Ray and Gamma-Ray Telescopes and Instruments for Astronomy. J. E. Truemper, H. D. Tananbaum (eds.), Proceedings of the SPIE, Volume 4851, 289

\bibitem{Whi03} White, N. E., \& Tananbaum, H. D. 2003, X-Ray and Gamma-Ray Telescopes and Instruments for Astronomy, J. E. Truemper, H. D. Tananbaum (eds.), Proceedings of the SPIE, Volume 4851, 293

\bibitem{Par04} Parmar, A., Hasinger, G., \& Turner, M. 2004, 34th COSPAR Scientific Assembly, The Second World Space Congress, held 10-19 October, 2002 in Houston, TX, USA., p.2368 


\end{thebibliography}
\end{document}